\documentclass[twocolumn]{article}
\usepackage[utf8]{inputenc}
\usepackage{textcomp,marvosym}
\usepackage[table,xcdraw]{xcolor}
\usepackage{sectsty}
\allsectionsfont{\sffamily}
\usepackage{abstract}

\usepackage{arydshln}
\usepackage{url}
\usepackage{listings}
\usepackage{hyperref}
\usepackage{float}
\usepackage{enumitem}
\setlist[enumerate]{noitemsep}
\usepackage{nameref,hyperref}
\usepackage{textgreek}
\usepackage{graphicx}
\usepackage[left=0.6in,right=0.6in,top=0.6in,bottom=1in]{geometry}
\usepackage{authblk}
\usepackage[backend=biber,style=nature,sorting=none]{biblatex}
\providecommand{\keywords}[1]{\textbf{\textit{Keywords --}} #1}

\begin{document}

\title{\sffamily Cultural transmission modes of music sampling traditions remain stable despite delocalization in the digital age\rmfamily}

\date{}

\author[a,b,1]{\normalsize Mason Youngblood}

\affil[a]{\scriptsize Department of Psychology, The Graduate Center, City University of New York, New York, NY, USA}
\affil[b]{\scriptsize Department of Biology, Queens College, City University of New York, Flushing, NY, USA\newline\textsuperscript{1}myoungblood@gradcenter.cuny.edu}

\twocolumn[
\begin{@twocolumnfalse}

\maketitle

\vspace*{-20pt}
\begin{abstract}
\vspace*{-10pt}

Music sampling is a common practice among hip-hop and electronic producers that has played a critical role in the development of particular subgenres. Artists preferentially sample drum breaks, and previous studies have suggested that these may be culturally transmitted. With the advent of digital sampling technologies and social media the modes of cultural transmission may have shifted, and music communities may have become decoupled from geography. The aim of the current study was to determine whether drum breaks are culturally transmitted through musical collaboration networks, and to identify the factors driving the evolution of these networks. Using network-based diffusion analysis we found strong evidence for the cultural transmission of drum breaks via collaboration between artists, and identified several demographic variables that bias transmission. Additionally, using network evolution methods we found evidence that the structure of the collaboration network is no longer biased by geographic proximity after the year 2000, and that gender disparity has relaxed over the same period. Despite the delocalization of communities by the internet, collaboration remains a key transmission mode of music sampling traditions. The results of this study provide valuable insight into how demographic biases shape cultural transmission in complex networks, and how the evolution of these networks has shifted in the digital age.

\\\\

\end{abstract}
\end{@twocolumnfalse}]

\section*{Introduction}

Music sampling, or the use of previously-recorded material in a new composition, is a nearly ubiquitous practice among hip-hop and electronic producers. The usage of drum breaks, or percussion-heavy sequences, ripped from soul and funk records has played a particularly critical role in the development of certain subgenres. For example, ``Amen, Brother'', released by The Winstons in 1969, is widely regarded as the most sampled song of all time. Its iconic 4-bar drum break has been described as ``genre-constitutive'' \cite{Whelan2009}, and can be prominently heard in classic hip-hop and jungle releases by N.W.A and Shy FX \cite{Collins2007}. Due to the consistent usage of drum breaks in particular music communities and subgenres \cite{Whelan2009,Collins2007,Frane2017,Vakeva2010,Rodgers2003} some scholars have suggested that they may be culturally transmitted \cite{Bown2009}, which could occur as a direct result of collaboration between artists or as an indirect effect of community membership.

Before the digital age, artists may have depended upon collaborators for access to the physical source materials and expensive hardware required for sampling \cite{Lloyd2014}. In the 1990s, new technologies like compressed digital audio formats and digital audio workstations made sampling more accessible to a broader audience \cite{Bodiford2017}. Furthermore, the widespread availability of the internet and social media have delocalized communities \cite{DiMaggio2001}, and allowed global music ``scenes'' to form around shared interests beyond peer-to-peer file sharing \cite{Ebare2003,Alexandraki2009}. Individuals in online music communities now have access to the collective knowledge of other members \cite{Salavuo2006,Lazar2002}, and there is evidence that online communities play a key role in music discovery \cite{Garg2011}. Although musicians remain concentrated in historically important music cities (i.e. New York City and Los Angeles in the United States) \cite{Florida2010,Graham2016}, online music communities also make it possible for artists to establish collaborative relationships independently of geographic location \cite{Kruse2010}. If more accessible sampling technologies and access to collective knowledge have allowed artists to discover sample sources independently of collaboration \cite{Makelberge2012}, then the strength of cultural transmission via collaboration may have decreased over the last couple of decades. Similarly, if online music communities have created opportunities for interactions between potential collaborators, then geographic proximity may no longer structure musical collaboration networks.

Studies of the cultural evolution of music have primarily investigated diversity in musical performances \cite{Ellis2018} and traditions \cite{Savage2014}, macro-scale patterns and selective pressures in musical evolution \cite{Mauch2015,Percino2014,Savage2015,RodriguezZivic2013}, and the structure and evolution of consumer networks \cite{Garg2011,Schlitter2009,Monechi2017}. Although several diffusion chain experiments have addressed how cognitive biases shape musical traits during transmission \cite{Verhoef2012,Ravignani2017,Lumaca2017}, few studies have investigated the mechanisms of cultural transmission at the population level \cite{Rossman2008,Nakamura2018}. The practice of sampling drum breaks in hip-hop and electronic music is an ideal research model for cultural transmission because of (1) the remarkably high copy fidelity of sampled material, (2) the reliable documentation of sampling events, and (3) the availability of high-resolution collaboration and demographic data for the artists involved. Exhaustive online datasets of sample usage and collaboration make it possible to reconstruct networks of artists and track the diffusion of particular drum breaks from the early 1980s to today. Furthermore, the technological changes that have occurred over the same time period provide a natural experiment for how the digital age has impacted cultural transmission more broadly \cite{Acerbi2016}.

The aim of the current study was to determine whether drum breaks are culturally transmitted through musical collaboration networks, and to identify the factors driving the evolution of these networks. We hypothesized that (1) drum breaks are culturally transmitted through musical collaboration networks, and that (2) the strength of cultural transmission via collaboration would decrease after the year 2000. For clarification, the alternative to the first hypothesis is cultural transmission occurring outside of collaborative relationships (i.e. independent sample discovery via ``crate-digging'' in record stores or online). Previous studies have investigated similar questions using diffusion curve analysis \cite{Rossman2008}, but the validity of inferring transmission mechanisms from cumulative acquisition data has been called into question \cite{Laland2003}. Instead, we applied network-based diffusion analysis (NBDA), a recently developed statistical method for determining whether network structure biases the emergence of a novel behavior in a population \cite{Franz2009}. As NBDA is most useful in identifying social learning, an ability that is assumed to be present in humans, it has been primarily applied to non-human animal models such as birds, whales, and primates \cite{Aplin2014,Allen2013,Hobaiter2014}, but the ability to incorporate individual-level variables to nodes makes it uniquely suited to determining what factors bias diffusion more generally. Additionally, we hypothesized that (3) collaboration probability would be decoupled from geographic proximity after the year 2000. To investigate this we applied separable temporal exponential random graph modeling (STERGM), a dynamic extension of ERGM for determining the variables that bias network evolution \cite{Krivitsky2010}.

\section*{Methods}

All data used in the current study were collected in September of 2018, in compliance with the terms and conditions of each database. For the primary analysis, the three most heavily sampled drum breaks of all time, ``Amen, Brother'' by The Winstons, ``Think (About It)'' by Lyn Collins, and ``Funky Drummer'' by James Brown, were identified using WhoSampled\footnote{\url{https://www.whosampled.com/}}. The release year and credits for each song listed as having sampled each break were collected using data scraping. In order to avoid name disambiguation, only artists, producers, and remixers with active Discogs links and associated IDs were included in the dataset. In order to investigate potential shifts in transmission strength around 2000, the same method was used to collect data for the eight songs in the ``Most Sampled Tracks'' on WhoSampled that were released after 1990 (see \nameref{S1_Appendix}). One of these, ``I'm Good'' by YG, was excluded from the analysis because the sample is primarily used by a single artist. Each set of sampling events collected from WhoSampled was treated as a separate diffusion. All analyses were conducted in R (v 3.3.3).

Collaboration data were retrieved from Discogs\footnote{\url{https://www.discogs.com/}}, a crowdsourced database of music releases. All collaborative releases in the database were extracted and converted to a master list of pairwise collaborations. For each diffusion, pairwise collaborations including two artists in the dataset were used to construct collaboration networks, in which nodes correspond to artists and weighted links correspond to collaboration number. Although some indirect connections between artists were missing from these subnetworks, conducting the analysis with the full dataset was computationally prohibitive and incomplete networks have been routinely used for NBDA in the past \cite{Aplin2012,Allen2013,Aplin2014}.

Individual-level variables for artists included in each collaboration network were collected from MusicBrainz\footnote{\url{https://musicbrainz.org/}}, a crowdsourced database with more complete artist information than Discogs, and Spotify\footnote{\url{https://www.spotify.com/}}, one of the most popular music streaming services. Gender and geographic location were retrieved from the Musicbrainz API. Whenever it was available, the ``begin area'' of the artist, or the city in which they began their career, was used instead of their ``area'', or country of affiliation, to maximize geographic resolution. Longitudes and latitudes for each location, retrieved using the Data Science Toolkit and Google Maps, were used to calculate each artist's mean geographic distance from other individuals. Albunack\footnote{\url{http://www.albunack.net/}}, an online tool which draws from both Musicbrainz and Discogs, was used to convert IDs between the two databases. Popularity and followers were retrieved using the Spotify API. An artist's popularity, a proprietary index of streaming count that ranges between 0 and 100, is a better indicator of their long-term success because it is calculated across their entire discography. Followers is a better indicator of current success because it reflects user engagement with artists who are currently more active on the platform. Discogs IDs are incompatible with the Spotify API, so artist names were URL-encoded and used as text search terms.

In order to identify whether social transmission between collaborators played a role in sample acquisition, order of acquisition diffusion analysis (OADA) was conducted using the R script for NBDA (v 1.2.13) provided on the Laland lab's website\footnote{\url{https://lalandlab.st-andrews.ac.uk/freeware/}}. OADA uses the order in which individuals acquire a trait to determine whether its diffusion is biased by the structure of the social network\cite{Hoppitt2010}. OADA was utilized instead of time of acquisition diffusion analysis (TADA) because it makes no assumptions about the baseline rate of acquisition \cite{Franz2009}. For each artist, order of acquisition was determined by the year that they first used the sample in their music. Sampling events from the same year were given the same order. Gender, popularity, followers, and mean distance were included as predictor variables. For gender, females were coded as -1, males were coded as 1, and individuals with other identities or missing values were coded as 0. For popularity, followers, and mean distance each variable was centered around zero. Asocial, additive, and multiplicative models were fit to all three diffusions collectively with every possible combination of individual-level variables. Standard information theoretic approaches were used to rank the models according to Akaike's Information Criterion corrected for sample size (AIC\textsubscript{c}). Models with a $\Delta$AIC\textsubscript{c} $<$ 2 were considered to have the best fit \cite{Burnham2002}. The best fitting model with the most individual-level variables was run separately to assess the effects of each variable on social transmission. Effect sizes were calculated according to \cite{Allen2013}. An additional OADA was conducted using the seven diffusions from after 1990. Individual-level variables were excluded due to insufficient demographic data. An additive model was fit to the OADA, and separate social transmission parameters were calculated for each diffusion to identify differences in transmission strength. Additive and multiplicative models give identical results in the absence of individual-level variables, so no model comparison was necessary.

In order to assess the effects of individual-level variables on network evolution, STERGM was conducted using statnet (v 2016.9), an R package for network analysis and simulation. STERGM is a dynamic social network method that models the formation and dissolution of links over time\cite{Krivitsky2010}. Collaboration events involving artists from each diffusion were combined to construct static collaboration subnetworks for each year between 1984 and 2017, which were then converted into an undirected, unweighted dynamic network. Early years not continuous with the rest of the event data (i.e. 1978 and 1981) were excluded from the dynamic network. In order to determine whether the variables biasing network structure have changed over time, the analysis was conducted separately with the data from 1984-1999 and 2000-2017. For each time period a set of STERGM models with every possible combination of individual-level variables were fit to the dynamic network using conditional maximum likelihood estimation (CMLE). Although STERGM can be used to separately model both the formation and dissolution of links, this analysis was restricted to the former. Gender, popularity, and followers were included to investigate homophily, while mean distance was included to assess its effect on link formation. As STERGM cannot be run with missing covariates, NA values in popularity (6.39\%), followers (6.39\%), and mean distance (38.49\%) were imputed using the random forest method. The models from each period were ranked according to AIC, and the best fitting models ($\Delta$AIC $<$ 2) with the most individual-level variables were run separately to assess the effects of each variable on network evolution.

\section*{Results}

The three most heavily sampled drum breaks of all time were collectively sampled 6530 times (\textit{n}\textsubscript{1} = 2966, \textit{n}\textsubscript{2} = 2099, \textit{n}\textsubscript{3} = 1465). 4462 (68.33\%) of these sampling events were associated with valid Discogs IDs, corresponding to 2432 unique artists (F: \textit{n} = 143, 5.88\%; M: \textit{n} = 1342, 55.18\%; Other or NA: \textit{n} = 947, 38.94\%), and included in the primary OADA and STERGM. The eight samples released after 1990 were collectively sampled 1752 times (\textit{n}\textsubscript{1} = 284, \textit{n}\textsubscript{2} = 260, \textit{n}\textsubscript{3} = 248, \textit{n}\textsubscript{4} = 198, \textit{n}\textsubscript{5} = 194, \textit{n}\textsubscript{6} = 193, \textit{n}\textsubscript{7} = 192, \textit{n}\textsubscript{8} = 182). 1305 (74.53\%) of these sampling events were associated with valid Discogs IDs, corresponding to 1270 unique artists, and included in the additional OADA.

\subsection*{NBDA}

\begin{figure*}
\centering
\includegraphics[width=0.8\linewidth]{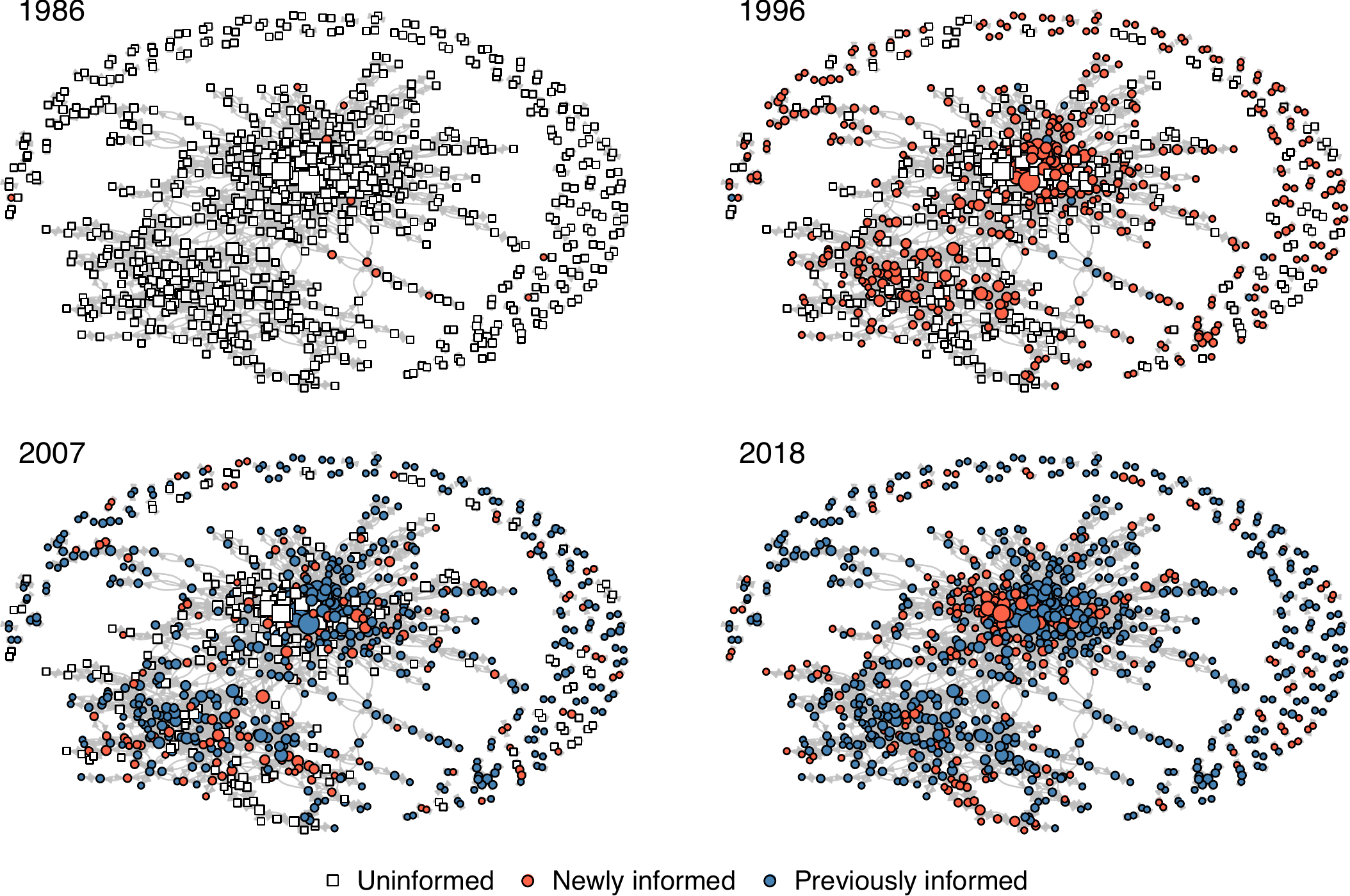}
\caption{The diffusion of all three drum breaks through the combined collaboration network. At each time point individuals who have not yet used one of the drum breaks (informed) are shown in white, individuals who first used one of the drum breaks in a previous time step (previously informed) are shown in blue, and individuals who first used one of the drum breaks in the current time step (newly informed) are shown in red.}
\label{diffusion_network}
\end{figure*}

The best fitting model from the primary OADA, which was multiplicative and included all four individual-level variables, can be seen in Table \ref{nbda_results}. In support of our first hypothesis, a likelihood ratio test found strong evidence for social transmission over asocial learning ($\Delta$AIC\textsubscript{c} = 141; \textit{p} $<$ 0.001). Based on the effect sizes, transmission appears to be more likely among females (\textit{p} $<$ 0.01) and less likely among artists who are more popular (\textit{p} $<$ 0.001) and have more followers (\textit{p} $<$ 0.001). Mean distance is not a significant predictor of transmission (\textit{p} = 0.89). The diffusion network and diffusion curve for all three drum breaks included in the primary OADA are shown in Figures \ref{diffusion_network} and \ref{diffusion_curve}, respectively. All other models fit to the primary OADA can be found in the supporting information.

\renewcommand{\arraystretch}{1.2}
\begin{table}[]
\centering
\scalebox{0.9}{\begin{tabular}{rccc}
\multicolumn{4}{l}{\cellcolor[HTML]{000000}{\color[HTML]{FFFFFF} Multiplicative Model - Order of Acquisition}}\\
\multicolumn{1}{l}{} & Estimate & Effect size & \textit{p}\\\hdashline
Gender & -0.11 & 0.81 & $<$ 0.01\\
Popularity & -0.013 & 0.86 & $<$ 0.001\\
Followers & -9.6E-8 & 0.92 & $<$ 0.001\\
Mean distance & -1.8E-9 & 1 & 0.89\\
\multicolumn{4}{l}{\cellcolor[HTML]{000000}{\color[HTML]{FFFFFF} Likelihood Ratio Test}}\\
\multicolumn{2}{l}{} & AIC\textsubscript{c} & \textit{p}\\\hline
\multicolumn{2}{r}{With social transmission} & 14719 & $<$ 0.001\\
\multicolumn{2}{r}{Without social transmission} & 14860 & \\\hline
\end{tabular}}
\caption{The results of the multiplicative model for the OADA including all individual-level variables. The top panel shows the model estimate, effect size, and \textit{p}-value for each individual-level variable. The bottom panel shows the AIC\textsubscript{c} for the model with and without social transmission and the \textit{p}-value from the likelihood ratio test.}
\label{nbda_results}
\end{table}

The results of the additional OADA, conducted using the seven diffusions from after 1990, can be found in the supporting information. A likelihood ratio test found strong evidence for social transmission overall ($\Delta$AIC\textsubscript{c} = 88; \textit{p} $<$ 0.001). Contrary to our second hypothesis, linear regression found no significant relationships between either mean year of diffusion and social transmission estimate (R\textsuperscript{2} = 0.20, \textit{p} = 0.31) or median year of diffusion and social transmission estimate (R\textsuperscript{2} = 0.17, \textit{p} = 0.36) (see Figure \ref{m_m_estimates}).

\subsection*{STERGM}

For both time periods the second best fitting STERGM models ($\Delta$AIC $<$ 2) included all four individual-level variables, the results of which can be seen in Table \ref{stergm_results}. All other models, including those assuming different transition years, can be found in the supporting information. Across both periods there appears to be homophily based on popularity (\textit{p} $<$ 0.001) and gender (M: \textit{p} $<$ 0.001; F: \textit{p}s $<$ 0.05). In support of our third hypothesis, mean distance negatively predicts link formation only before 2000 (\textit{p} $<$ 0.001). Additionally, there is a heterophilic effect of followers only after 2000 (\textit{p} $<$ 0.001). Based on the effect sizes, there has been a nearly three-fold decrease in the strength of homophily among females. Conversely, the strengh of homophily by popularity has actually increased since 2000. Linear regression found significant positive relationships between both popularity and number of collaborations (R\textsuperscript{2} = 0.048, \textit{p} $<$ 0.001) and followers and number of collaborations (R\textsuperscript{2} = 0.090, \textit{p} $<$ 0.001) (see Figure \ref{collab_corr}).

A goodness-of-fit analysis was conducted by generating simulated networks (n = 100) from the parameters of the best fitting model and comparing them to the observed network statistics \cite{Hunter2008}. For both time periods, the global statistics (i.e. gender, popularity, followers, mean distance) from the simulated networks were not significantly different from those observed, indicating that both models are good fits for the variables in question. Structural statistics (i.e. degree, edgewise shared partner, minimum geodesic distance) from the simulated networks were significantly different from those observed, indicating that both models are not good fits for the structural properties of the network. The results of this analysis can be found in the supporting information.

\renewcommand{\arraystretch}{1.2}
\begin{table}[]
\centering
\scalebox{0.9}{\begin{tabular}{rcccc}
\rowcolor[HTML]{000000}\multicolumn{1}{l}{\color[HTML]{FFFFFF} STERGM} & \multicolumn{2}{c}{\color[HTML]{FFFFFF} 1984-1999} & \multicolumn{2}{c}{\color[HTML]{FFFFFF} 2000-2017}\\
\multicolumn{1}{l}{} & Effect size & \textit{p} & Effect size & \textit{p}\\\hdashline
Gender (F) & 6.86 & $<$ 0.001 & 2.23 & $<$ 0.05\\
Gender (M) & 1.70 & $<$ 0.001 & 2.43 & $<$ 0.001\\
Popularity & 0.84 & $<$ 0.001 & 0.54 & $<$ 0.001\\
Followers & 1.02 & 0.64 & 1.96 & $<$ 0.001\\
Mean distance & 0.87 & $<$ 0.001 & 0.99 & 0.82 \\\hline
\end{tabular}}
\caption{The results of the STERGM analyses for before and after 2000. The table shows the effect size and \textit{p}-value for gender, popularity, followers, and mean distance during each time period.}
\label{stergm_results}
\end{table}

\section*{Discussion}

Using high-resolution collaboration and longitudinal diffusion data, we have provided the first quantitative evidence that music samples are culturally transmitted via collaboration between artists. Additionally, in support of the widespread assertion that the internet has delocalized artist communities, we have found evidence that geographic proximity no longer biases the structure of musical collaboration networks after the year 2000. Given that the strength of transmission has not weakened over the same time period, this finding indicates that collaboration remains a key cultural transmission mode for music sampling traditions. This result supports the idea that the internet has enhanced rather than disrupted existing social interactions \cite{DiMaggio2001}.

Gender appears to play a key role in both network structure and cultural transmission. Across the entire time period, collaborations were more likely to occur between individuals of the same gender. Additionally, the probability of cultural transmission appears to be much higher for female artists. This effect could be a result of the much higher levels of homophily among women before 2000. Previous work has suggested that high levels of gender homophily are associated with gender disparity \cite{Glass2017,Crewe2018,Jadidi2018}, which is consistent with the historic marginalization of women in music production communities \cite{Ebare2003,Baker2008,Whelan2009}. Although the proportion of female artists in the entire dataset is extremely low ($\sim$6\%), the reduction in homophily among female artists after 2000 could be reflective of increasing inclusivity \cite{Smith2014}.

Artists with similar levels of popularity were also more likely to collaborate with each other. The increase in homophily by popularity after 2000 could be the result of an increase in skew, whereby fewer artists take up a greater proportion of the music charts \cite{Ordanini2016}. In addition, the probability of cultural transmission appears to be higher among less popular artists, even though they are slightly less collaborative. This effect could be linked to cultural norms within ``underground'' music production communities. In these communities, collective cultural production is sometimes prioritized over individual recognition \cite{Thornton1995,Hesmondhalgh1998}. This principle is best demonstrated by the historic popularity of the white-label release format, where singles are pressed to blank vinyl and distributed without artist information \cite{Thornton1995,Hesmondhalgh1998}. In more extreme cases, individual artists who experience some level of mainstream success or press coverage risk losing credibility, and may even be perceived as undermining the integrity of their music community \cite{Hesmondhalgh1998,Noys1995}. Concerns about credibility could cause individuals to selectively copy less popular artists or utilize more rare samples (i.e. De La Soul's refusal to sample James Brown and George Clinton because of their use by other popular groups \cite{Lena2004}). Future research should investigate whether the ``high prestige attached to obscurity'' \cite{Hesmondhalgh1998} in these communities may be driving a model-based bias for samples used by less popular artists or a frequency-based bias for samples that are more rare in the population \cite{Boyd1985}. A frequency-based novelty bias was recently identified in Western classical music using agent-based modeling \cite{Nakamura2018}, and similar methods could be utilized for sampling.

Similarly to popularity, the number of followers an individual has negatively predicts transmission probability. However, artists with similar numbers of followers were actually less likely to collaborate with each other after 2000. This result could be due to the fact that followers is a better indicator of current popularity, but has lower resolution further back in time. Newer artists with inflated follower counts who collaborate with older, historically-important artists with lower follower counts may still be expressing homophily based on overall popularity.

There are several limitations to this study that should be highlighted. Firstly, Discogs primarily documents official releases, which means that more recent releases on streaming sites like Soundcloud are not well-represented. In combination with the exclusion of artists without Discogs IDs, this indicates that less prominent artists may be underrepresented in the dataset. Fortunately, social networks are fairly robust to missing data, especially when networks are large, centralized, and disproportionately include central nodes \cite{Smith2013,Smith2017}. Additionally, simulation studies evaluating the robustness of NBDA indicate that it performs well under fairly high levels of sampling error and bias \cite{Franz2009,Whalen2016,Hoppitt2017,Wild2018}, risks that are mitigated by the fact that the network was reconstructed from published collaborations rather than temporal co-occurence data \cite{Franz2010}. Secondly, the time lag inherent in the user editing of WhoSampled means that older transmission records are more complete. Algorithms for sample-detection \cite{Hockman2015} may allow researchers to reconstruct full transmission records in the future, but these approaches are not yet publicly available. Lastly, MusicBrainz and Spotify had incomplete demographic data for some artists (i.e. gender and geographic location), which may have introduced bias into our model estimates.

The results of this study provide valuable insight into how demographic variables, particularly gender and popularity, have biased both cultural transmission and the evolution of collaboration networks going into the digital age. In addition, we provide evidence that collaboration remains a key transmission mode of music sampling traditions despite the delocalization of communities by the internet. Future research should investigate whether decreased homophily among females is actually linked to greater inclusivity in the music industry (e.g. booking rates, financial compensation, media coverage), as well as whether the inverse effect of popularity on cultural transmission probability is a result of a model-based bias for obscurity or a frequency-based bias for novelty.

\section*{\large Acknowledgments}

I would like to thank David Lahti and Carolyn Pytte, as well as all members of the Lahti lab, for their valuable conceptual and analytical feedback.

\section*{\large Data Availability Statement}

All R scripts and data used in the study are available in the Harvard Dataverse repository: \url{https://doi.org/10.7910/DVN/Q02JJQ}.

\renewcommand*{\bibfont}{\scriptsize}
\printbibliography[title=\large References]

\newpage
\onecolumn

\begin{center}
\sffamily\LARGE Supporting information\rmfamily
\end{center}

\setcounter{figure}{0}
\setcounter{subsection}{0}
\renewcommand{\thetable}{S\arabic{table}}
\renewcommand{\thefigure}{S\arabic{figure}}
\setlength\parindent{0pt}

\section*{NBDA}

\subsection{Primary OADA}

The results of the multiplicative NBDA model fit to the primary OADA with all four individual-level variables are shown below.

\smallskip
\lstset{basicstyle=\scriptsize,xleftmargin=.25in}
\begin{lstlisting}
Summary of Multiplicative Social Transmission Model
Order of acquisition data
Unbounded parameterisation.

Coefficients:
                           Estimate   Bounded           se          z            p
Social transmission 1  1.334607e-01 0.1177462           NA         NA           NA
gender                -1.050416e-01        NA 4.017186e-02 -2.6148064 8.927804e-03
popularity            -1.335417e-02        NA 1.465003e-03 -9.1154550 0.000000e+00
followers             -9.611204e-08        NA 1.956053e-08 -4.9135695 8.943303e-07
meandist              -1.880816e-09        NA 1.361273e-08 -0.1381659 8.901093e-01


Likelihood Ratio Test for Social Transmission:

Null model includes all other specified variables
Social transmission and asocial learning assumed to combine multiplicatively

                            Df LogLik   AIC  AICc     LR  p
With Social Transmission     5 7354.4 14719 14719 143.09  0
Without Social Transmission  4 7425.9 14860 14860          
\end{lstlisting}
\smallskip

The results of all NBDA models fit to the primary OADA. In the ``Additive?'' column TRUE means the model was additive, FALSE means the model was multiplicative, and NA means the model was asocial. In the ``ILVs'', or individual-level variables, column the numbers correspond to the variables included in the model (1: gender; 2: popularity; 3: followers; 4: mean distance).

\smallskip
\begin{lstlisting}
Additive?	ILVs	Social?	AICc			deltaAICc
FALSE		1 2 3 4	social	14718.7305939155	0
FALSE		2 3 4	social	14723.2581626229	4.53
FALSE		1 2 4	social	14749.7210576547	30.99
FALSE		2 4	social	14755.0075144196	36.28
FALSE		1 3 4	social	14795.6248094134	76.89
FALSE		3 4	social	14796.5052668663	77.77
FALSE		4	social	15956.6213244064	1237.89
FALSE		1 4	social	15957.1346280933	1238.4
FALSE		1 2 3	social	32229.4897515171	17510.76
FALSE		1 2	social	32260.0747802103	17541.34
FALSE		2 3	social	32267.3112186711	17548.58
FALSE		2	social	32298.2544538564	17579.52
FALSE		1 3	social	32356.8941850595	17638.16
FALSE		3	social	32389.9528866174	17671.22
FALSE		1	social	43609.4787479712	28890.75
TRUE		1	social	43612.45988987		28893.73
NA		0	social	43662.7306996825	28944
NA		1	asocial	43779.4569488198	29060.73
NA		0	asocial	43815.27922266		29096.55
TRUE		2	social	Inf			Inf
NA		2	asocial	Inf			Inf
TRUE		1 2	social	Inf			Inf
NA		1 2	asocial	Inf			Inf
TRUE		3	social	Inf			Inf
NA		3	asocial	Inf			Inf
TRUE		1 3	social	Inf			Inf
NA		1 3	asocial	Inf			Inf
TRUE		2 3	social	Inf			Inf
NA		2 3	asocial	Inf			Inf
TRUE		1 2 3	social	Inf			Inf
NA		1 2 3	asocial	Inf			Inf
TRUE		4	social	Inf			Inf
NA		4	asocial	Inf			Inf
TRUE		1 4	social	Inf			Inf
NA		1 4	asocial	Inf			Inf
TRUE		2 4	social	Inf			Inf
NA		2 4	asocial	Inf			Inf
TRUE		1 2 4	social	Inf			Inf
NA		1 2 4	asocial	Inf			Inf
TRUE		3 4	social	Inf			Inf
NA		3 4	asocial	Inf			Inf
TRUE		1 3 4	social	Inf			Inf
NA		1 3 4	asocial	Inf			Inf
TRUE		2 3 4	social	Inf			Inf
NA		2 3 4	asocial	Inf			Inf
TRUE		1 2 3 4	social	Inf			Inf
NA		1 2 3 4	asocial	Inf			Inf
\end{lstlisting}
\smallskip

\begin{figure}[H]
\centering
\includegraphics[scale=0.50]{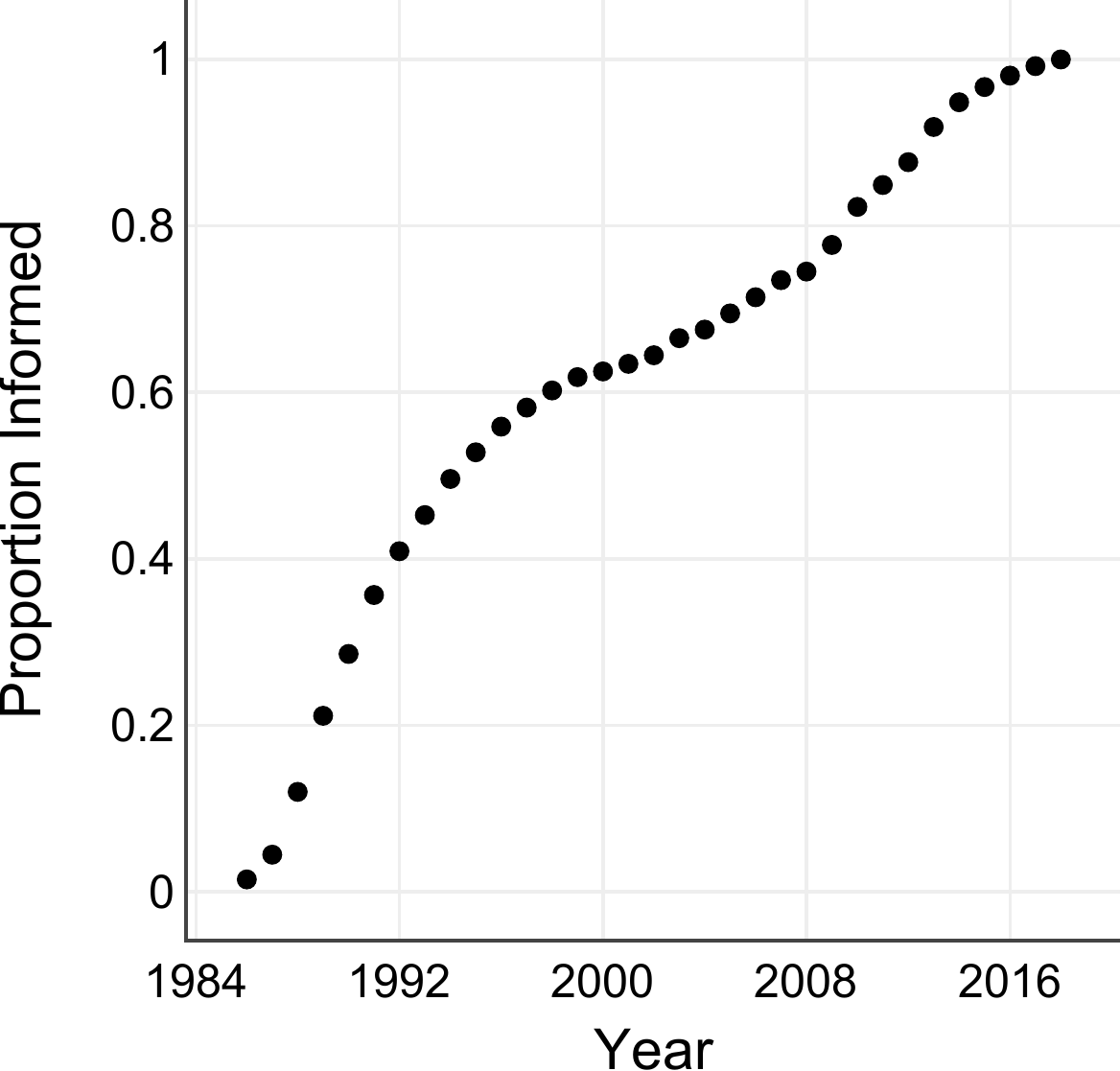}
\caption{The combined diffusion curve for all three drum breaks included in the primary OADA. The proportion of informed individuals is on the \textit{y}-axis, and the year is on the \textit{x}-axis. Although recent research suggests that inferring acquisition modes from diffusion curves is unreliable, it appears that the curve may have the \textit{S}-shape indicative of social transmission prior to the early-2000s.}
\label{diffusion_curve}
\end{figure}

\subsection{Additional OADA}

The eight songs in the ``Most Sampled Tracks'' on WhoSampled that were released after 1990 are shown below. The fifth song, ``I'm Good'' by YG, was excluded from the additional OADA because it is a producer tag used by a single artist.

\begin{enumerate}
\item ``Crash Goes Love (Yell Apella)'' by Loleatta Holloway (1992)
\item ``Shook Ones Part II'' by Mobb Deep (1994)
\item ``C.R.E.A.M.'' by Wu-Tang Clan (1993)
\item ``Sound of Da Police'' by KRS-One (1993)
\item ``I'm Good'' by YG (2011) [excluded producer tag]
\item ``Juicy'' by The Notorious B.I.G. (1994)
\item ``Sniper'' by DJ Trace and Pete Parsons (1999)
\item ``Who U Wit?'' by Lil Jon and The East Side Boyz (1997)
\end{enumerate}

The results of the additive NBDA model fit to the additional OADA are shown below. Remember that the fifth song was excluded, so the transmission estimates for five, six, and seven here are actually for six, seven, and eight.

\smallskip
\begin{lstlisting}
Summary of Additive Social Transmission Model
Order of acquisition data
Unbounded parameterisation

Coefficients
                        Estimate    Bounded
Social transmission 1 0.13558602 0.11939740
Social transmission 2 0.28805974 0.22363849
Social transmission 3 0.05184340 0.04928814
Social transmission 4 0.60154771 0.37560399
Social transmission 5 0.06816578 0.06381573
Social transmission 6 0.07600555 0.07063677
Social transmission 7 0.01547410 0.01523830


Likelihood Ratio Test for Social Transmission:

Null model includes all other specified variables
Social transmission and asocial learning assumed to combine additively

                            Df LogLik   AIC  AICc     LR  p
With Social Transmission     7   6450 12914 12914 101.99  0
Without Social Transmission  0   6501 13002 13002          
\end{lstlisting}
\smallskip

\begin{figure}[H]
\centering
\includegraphics[scale=0.50]{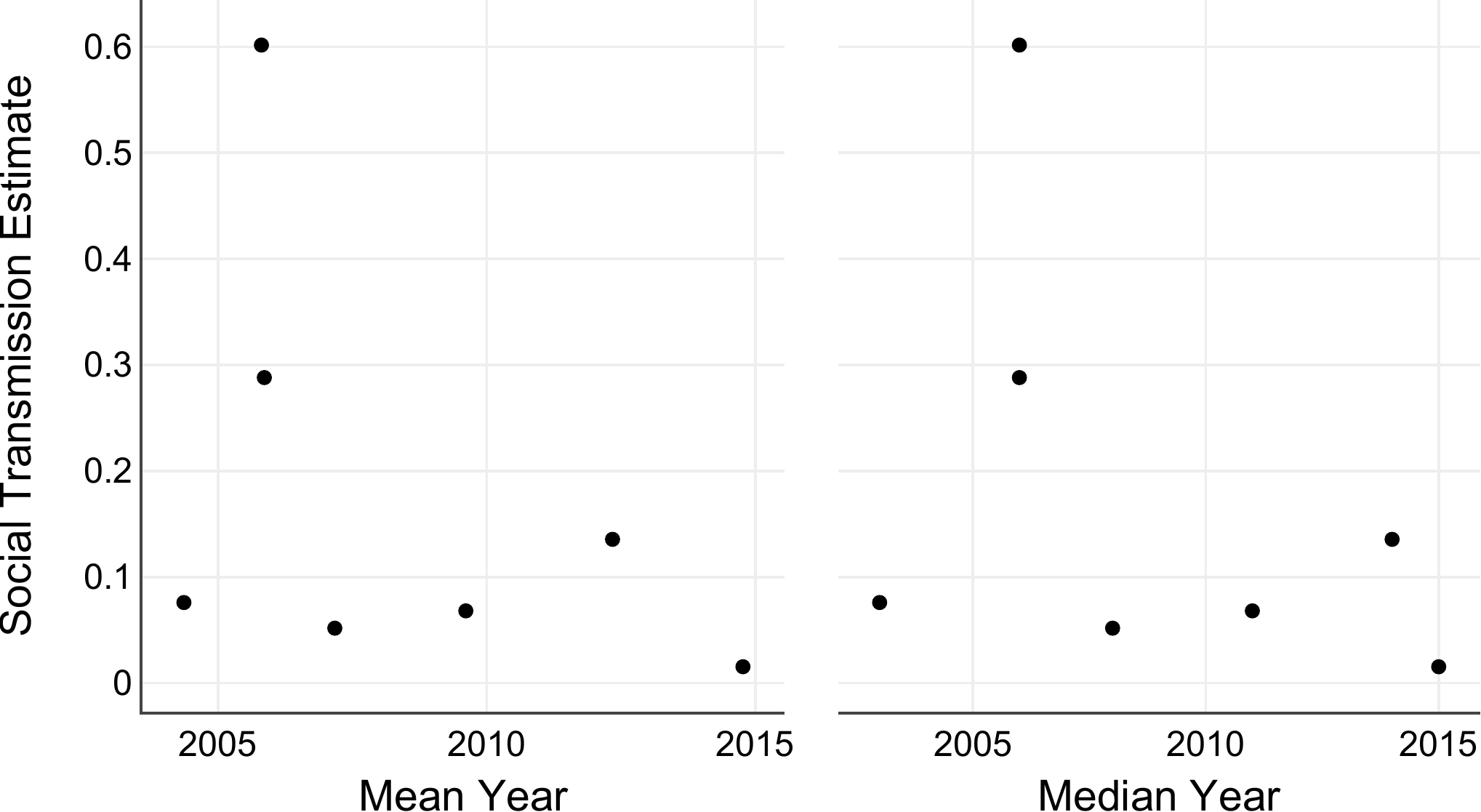}
\caption{The relationship between diffusion years and transmission strengths for all seven diffusions included in the additional OADA. The mean (left) and median (right) years of diffusion are on the \textit{x}-axis, and the social transmission estimates from the additive model are on the \textit{y}-axis. Linear regression found no significant relationships between either mean year of diffusion and social transmission estimate (R\textsuperscript{2} = 0.20, \textit{p} = 0.31) or median year of diffusion and social transmission estimate (R\textsuperscript{2} = 0.17, \textit{p} = 0.36).}
\label{m_m_estimates}
\end{figure}

\section*{STERGM}

The results of all formation models of the STERGM fit to the data from 1984-1999. In the ``ILVs'', or individual-level variables, column the numbers correspond to the variables included in the model (1: gender; 2: popularity; 3: followers; 4: mean distance).

\smallskip
\begin{lstlisting}
   ILVs      AIC deltaAIC
  1 2 4 7676.380  0.00000
1 2 3 4 7678.174  1.79348
    1 2 7693.425 17.04473
  1 2 3 7695.203 18.82240
    1 4 7702.003 25.62300
  1 3 4 7702.160 25.77921
    2 4 7710.403 34.02256
  2 3 4 7711.909 35.52898
      1 7718.147 41.76695
    1 3 7718.390 42.00953
      2 7727.194 50.81323
    2 3 7728.701 52.32017
      4 7737.373 60.99249
    3 4 7738.094 61.71319
      0 7753.228 76.84742
      3 7753.985 77.60457
\end{lstlisting}
\smallskip

The results of the best-fitting formation model of the STERGM with the most individual-level variables fit to the data from 1984-1999.

\smallskip
\begin{lstlisting}
==========================
Summary of model fit
==========================

Formula:   y.form ~ edges + nodecov("meandist") + absdiff("popularity") + 
    absdiff("followers") + nodematch("gender", diff = TRUE)
<environment: 0x1cb31da8>

Iterations:  11 out of 20 

Monte Carlo MLE Results:
                      Estimate Std. Error MCMC % z value Pr(>|z|)    
edges               -9.382e+00  1.089e-01      0 -86.175  < 1e-04 ***
nodecov.meandist    -1.292e-07  3.371e-08      0  -3.833 0.000126 ***
absdiff.popularity  -1.618e-02  3.284e-03      0  -4.927  < 1e-04 ***
absdiff.followers    1.194e-08  2.579e-08      0   0.463 0.643396    
nodematch.gender.-1  1.926e+00  3.293e-01      0   5.851  < 1e-04 ***
nodematch.gender.0   3.478e-01  1.965e-01      0   1.770 0.076773 .  
nodematch.gender.1   5.313e-01  1.117e-01      0   4.757  < 1e-04 ***
---
Signif. codes:  0 '***' 0.001 '**' 0.01 '*' 0.05 '.' 0.1 ' ' 1

     Null Deviance: 6349494  on 4580192  degrees of freedom
 Residual Deviance:    7664  on 4580185  degrees of freedom
 
AIC: 7678    BIC: 7772    (Smaller is better.) 
\end{lstlisting}
\smallskip

The results of the goodness-of-fit analysis of the formation model of the STERGM with the most individual-level variables fit to the data from 1984-1999 are below.

\smallskip
\begin{lstlisting}
Goodness-of-fit for degree 

     obs   min     mean   max MC p-value
0  10698 10388 10484.50 10553       0.00
1    801  1032  1101.97  1200       0.00
2    148    89   104.41   126       0.00
3     43    16    20.71    29       0.00
4     15     5     8.01    11       0.00
5      0     0     0.38     2       1.00
6      6     0     0.95     2       0.00
7      1     0     0.07     1       0.14
8     12     4     6.55     7       0.00
9      1     0     0.42     3       0.62
10     1     0     0.03     1       0.06
11     2     0     0.91     1       0.00
12     2     0     1.00     2       0.14
13     0     0     0.08     1       1.00
14     0     0     0.01     1       1.00

Goodness-of-fit for edgewise shared partner 

     obs min   mean max MC p-value
esp0 533 608 645.34 700          0
esp1  92  43  43.01  44          0
esp2  42  19  19.01  20          0
esp3   6   3   3.00   3          0
esp7  71  36  36.00  36          0
esp8   1   0   0.00   0          0

Goodness-of-fit for minimum geodesic distance 

         obs      min        mean      max MC p-value
1        745      709      746.36      801       0.92
2        421      186      212.11      249       0.00
3        243       46       67.21      102       0.00
4        138       14       22.30       45       0.00
5         60        6        8.67       22       0.00
6         19        0        1.07        7       0.00
7          5        0        0.06        3       0.00
Inf 68788954 68789399 68789527.22 68789605       0.00

Goodness-of-fit for model statistics 

                              obs           min          mean           max MC p-value
edges                      745.00        709.00        746.36        801.00       0.92
nodecov.meandist    -298388294.47 -366967354.28 -300622250.67 -216526380.89       0.94
absdiff.popularity       16092.87      15023.48      16140.46      17764.91       0.94
absdiff.followers    631702714.34  538012952.78  636527414.86  770413121.94       1.00
nodematch.gender.-1         20.00         13.00         20.16         28.00       1.00
nodematch.gender.0          68.00         56.00         68.37         80.00       1.00
nodematch.gender.1         384.00        361.00        385.54        424.00       1.00
\end{lstlisting}
\smallskip

The results of all formation models of the STERGM fit to the data from 2000-2017. In the ``ILVs'', or individual-level variables, column the numbers correspond to the variables included in the model (1: gender; 2: popularity; 3: followers; 4: mean distance).

\smallskip
\begin{lstlisting}
   ILVs      AIC  deltaAIC
  1 2 3 14194.16   0.00000
1 2 3 4 14196.11   1.94688
    2 3 14365.84 171.67787
  2 3 4 14367.30 173.13943
    1 2 14409.22 215.06168
  1 2 4 14410.98 216.82145
    1 3 14563.71 369.55351
  1 3 4 14565.71 371.55212
      2 14599.55 405.39275
    2 4 14600.59 406.43284
      1 14636.97 442.80822
    1 4 14638.95 444.78763
      3 14748.97 554.80763
    3 4 14750.71 556.55346
      0 14836.01 641.84666
      4 14837.57 643.40921
\end{lstlisting}
\smallskip

The results of the best-fitting formation model of the STERGM with the most individual-level variables fit to the data from 2000-2017.

\smallskip
\begin{lstlisting}
==========================
Summary of model fit
==========================

Formula:   y.form ~ edges + nodecov("meandist") + absdiff("popularity") + 
    absdiff("followers") + nodematch("gender", diff = TRUE)
<environment: 0x1ad107308>

Iterations:  10 out of 20 

Monte Carlo MLE Results:
                      Estimate Std. Error MCMC %  z value Pr(>|z|)    
edges               -8.530e+00  7.639e-02      0 -111.660  < 1e-04 ***
nodecov.meandist    -3.907e-09  1.702e-08      0   -0.230  0.81847    
absdiff.popularity  -4.764e-02  2.799e-03      0  -17.022  < 1e-04 ***
absdiff.followers    1.726e-07  9.188e-09      0   18.786  < 1e-04 ***
nodematch.gender.-1  8.030e-01  3.605e-01      0    2.227  0.02592 *  
nodematch.gender.0  -7.953e-01  2.582e-01      0   -3.080  0.00207 ** 
nodematch.gender.1   8.877e-01  7.878e-02      0   11.269  < 1e-04 ***
---
Signif. codes:  0 '***' 0.001 '**' 0.01 '*' 0.05 '.' 0.1 ' ' 1

     Null Deviance: 7195553  on 5190494  degrees of freedom
 Residual Deviance:   14182  on 5190487  degrees of freedom
 
AIC: 14196    BIC: 14290    (Smaller is better.) 
\end{lstlisting}
\smallskip

The results of the goodness-of-fit analysis of the formation model of the STERGM with the most individual-level variables fit to the data from 2000-2017 are below.

\smallskip
\begin{lstlisting}
Goodness-of-fit for degree 

     obs   min     mean   max MC p-value
0  11179 10679 10811.81 10930       0.00
1   1507  1878  2005.54  2115       0.00
2    371   313   351.05   398       0.24
3    130    76    88.47   104       0.00
4     68    17    25.10    31       0.00
5     26     3     7.85    13       0.00
6      5     0     1.94     6       0.06
7      5     0     1.94     3       0.00
8      0     0     0.26     2       1.00
9      2     0     0.04     1       0.00
10     1     0     0.00     0       0.00

Goodness-of-fit for edgewise shared partner 

      obs  min    mean  max MC p-value
esp0 1176 1324 1379.71 1458          0
esp1  294  144  144.42  147          0
esp2   75   35   36.03   37          0
esp3   22   10   10.02   11          0

Goodness-of-fit for minimum geodesic distance 

         obs      min        mean      max MC p-value
1       1567     1514     1570.18     1648       0.90
2       1189      607      677.37      735       0.00
3       1107      320      373.55      436       0.00
4        950      177      225.67      286       0.00
5        658       83      123.00      177       0.00
6        380       37       69.22      125       0.00
7        182       15       39.00       82       0.00
8         79        3       19.09       47       0.00
9         16        1        8.26       34       0.22
10         1        0        2.59       14       1.00
11         0        0        0.58        7       1.00
12         0        0        0.12        4       1.00
13         0        0        0.02        1       1.00
Inf 88352442 88355141 88355462.35 88355763       0.00

Goodness-of-fit for model statistics 

                              obs           min          mean          max MC p-value
edges                     1567.00       1514.00       1570.18 1.648000e+03       0.90
nodecov.meandist     -52499302.28 -196776194.24  -55329463.86 1.010890e+08       0.96
absdiff.popularity       25911.58      24355.87      25912.42 2.714093e+04       0.98
absdiff.followers   2958736083.69 2685506866.45 2938426320.54 3.215706e+09       0.86
nodematch.gender.-1         16.00         10.00         16.04 2.400000e+01       1.00
nodematch.gender.0          31.00         23.00         31.33 4.200000e+01       1.00
nodematch.gender.1        1012.00        956.00       1014.21 1.069000e+03       0.94
\end{lstlisting}
\smallskip

\begin{figure}[H]
\centering
\includegraphics[scale=0.50]{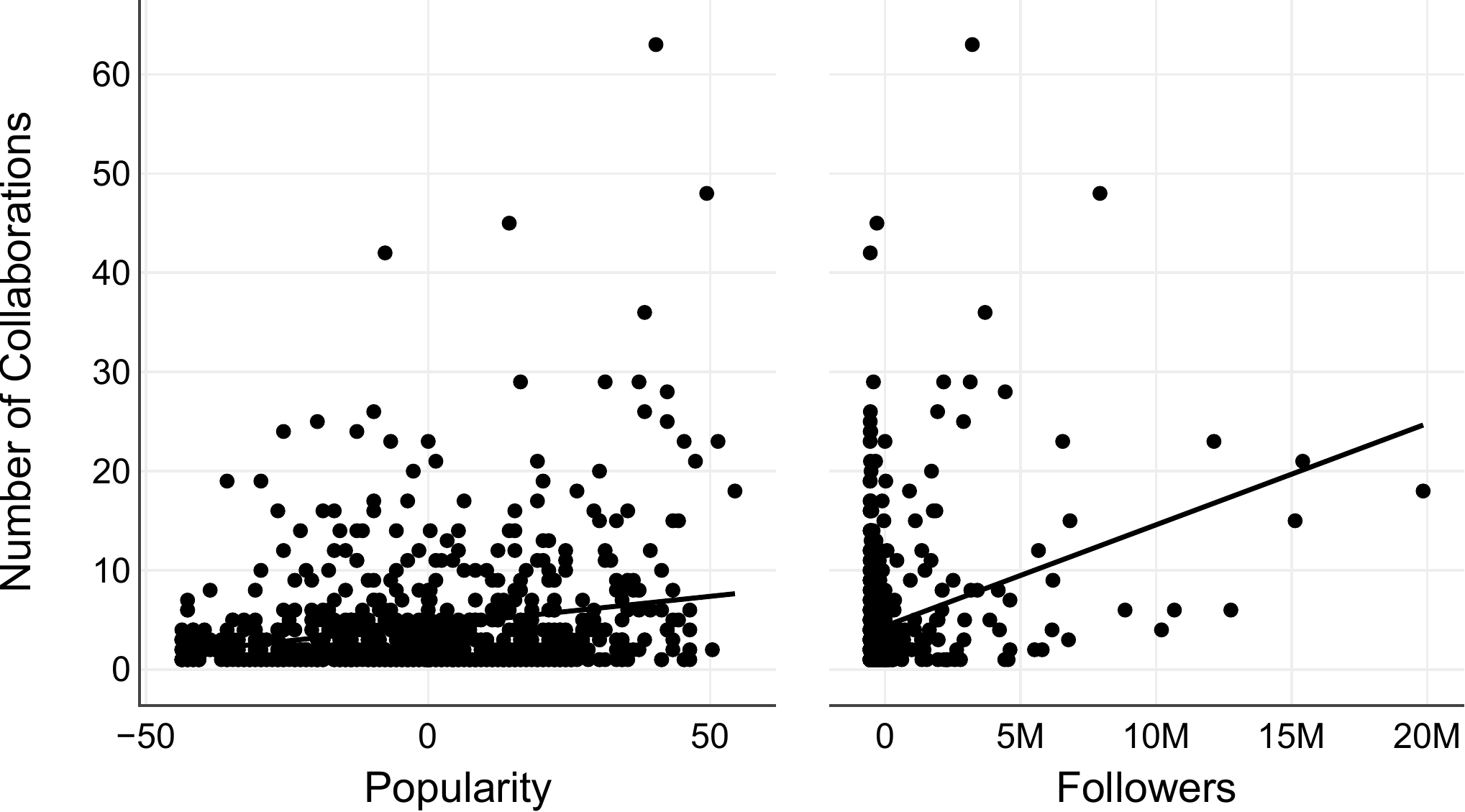}
\caption{The relationship between popularity and followers and the number of collaborations for each artist in the dataset. Popularity and followers are on the \textit{x}-axis, and number of collaborations is on the \textit{y}-axis. Linear regression found significant positive relationships between both popularity and number of collaborations (R\textsuperscript{2} = 0.048, \textit{p} $<$ 0.001) and followers and number of collaborations (R\textsuperscript{2} = 0.090, \textit{p} $<$ 0.001).}
\label{collab_corr}
\end{figure}

The results of the formation models of the STERGM with all individual-level variables assuming different transition years. The top and bottom tables show the results for before and after each transition year. The number of unique artists in each time period is included in the second row. Regardless of the transition year, mean distance and gender (F and M) had the same significance pattern and direction of effect observed in the main analysis. The results for popularity only varied from the main analysis in the first time period when the transition year was 1994 or 1996, which could be the result of lower sample sizes. The results for followers were consistent with the main analysis in the second time period, but fluctuated dramatically across transition years in the first time period.

\smallskip
\begin{lstlisting}
              Pre-1994         Pre-1996         Pre-1998         Pre-2000         
              n = 205          n = 286          n = 370          n = 450          
              Estimate p-value Estimate p-value Estimate p-value Estimate p-value 
Mean Distance -2.0e-07 9.0e-04 -1.0e-07 1.1e-02 -1.1e-07 3.2e-03 -1.3e-07 1.3e-04 
Popularity    -5.1e-03 2.9e-01 -5.9e-03 1.6e-01 -1.2e-02 1.5e-03 -1.6e-02 8.4e-07 
Followers     -2.3e-07 8.0e-03 -2.4e-07 1.9e-03 -8.3e-08 5.5e-02  1.2e-08 6.4e-01  
Gender (F)     1.5e+00 1.0e-02  1.3e+00 3.1e-02  1.3e+00 1.3e-02  1.9e+00 4.9e-09  
Gender (M)     5.8e-02 7.3e-01  2.6e-01 7.4e-02  4.7e-01 1.4e-04  5.3e-01 2.0e-06  

              Pre-2002         Pre-2004         Pre-2006        
              n = 520          n = 593          n = 659        
              Estimate p-value Estimate p-value Estimate p-value
              -1.1e-07 1.6e-04 -1.2e-07 1.2e-05 -1.1e-07 2.0e-05
              -1.8e-02 1.7e-09 -2.0e-02 9.9e-13 -2.2e-02 4.4e-17
               2.8e-08 2.0e-01  3.9e-08 4.3e-02  4.6e-08 8.0e-03
               1.8e+00 5.1e-09  1.7e+00 6.1e-08  1.5e+00 1.6e-06
               5.7e-01 2.0e-08  6.7e-01 5.8e-13  6.7e-01 7.5e-15
\end{lstlisting}
\smallskip

\smallskip
\begin{lstlisting}
              Post-1994        Post-1996        Post-1998        Post-2000        
              n = 876          n = 836          n = 781          n = 725          
              Estimate p-value Estimate p-value Estimate p-value Estimate p-value 
Mean Distance -2.0e-08 2.1e-01 -2.2e-08 1.7e-01 -9.3e-09 5.8e-01 -3.9e-09 8.2e-01 
Popularity    -4.3e-02 9.3e-72 -4.4e-02 2.4e-70 -4.7e-02 1.7e-68 -4.8e-02 5.7e-65 
Followers      1.6e-07 2.6e-74  1.6e-07 1.6e-77  1.7e-07 1.1e-75  1.7e-07 9.9e-79  
Gender (F)     1.3e+00 3.7e-07  1.3e+00 9.2e-07  8.3e-01 1.4e-02  8.0e-01 2.6e-02  
Gender (M)     8.9e-01 7.3e-37  9.0e-01 2.0e-35  8.6e-01 2.8e-30  8.9e-01 1.9e-29  

              Post-2002        Post-2004        Post-2006        
              n = 667          n = 607          n = 544        
              Estimate p-value Estimate p-value Estimate p-value
               1.2e-09 9.4e-01  1.7e-08 3.4e-01  2.2e-08 2.6e-01
              -4.9e-02 3.4e-61 -5.3e-02 2.4e-58 -5.5e-02 3.1e-52
               1.8e-07 3.2e-77  1.9e-07 1.9e-81  2.0e-07 1.2e-83
               8.7e-01 1.5e-02  9.7e-01 7.6e-03  1.2e+00 1.4e-03
               8.6e-01 1.5e-25  8.4e-01 4.7e-21  9.2e-01 5.6e-21
\end{lstlisting}
\smallskip

\end{document}